\documentclass[fleqn,usenatbib]{mnras}

\usepackage{newtxtext,newtxmath}

\usepackage[T1]{fontenc}

\DeclareRobustCommand{\VAN}[3]{#2}
\let\VANthebibliography\thebibliography
\def\thebibliography{\DeclareRobustCommand{\VAN}[3]{##3}\VANthebibliography}

\usepackage{longtable}
\usepackage{graphicx}	
\usepackage{amsmath}	

\title[Physical characterization of Gault]{(6478) Gault: Physical characterization of an active main-belt asteroid}

\author[M. Devog\`{e}le et al.]{
Maxime Devog\`{e}le,$^{1}$\thanks{E-mail: mdevogele@ucf.edu (MD)}
Marin Ferrais,$^{2}$
Emmanuel Jehin,$^{3}$
Nicholas Moskovitz,$^{4}$
Brian A. Skiff,$^{4}$\newauthor
Stephen E. Levine,$^{4,5}$
Annika Gustafsson,$^{6}$
Davide Farnocchia,$^{7}$
Marco Micheli,$^{8}$
Colin Snodgrass,$^{9}$\newauthor
Galin Borisov,$^{10,11}$
Jean Manfroid,$^{12}$
Youssef Moulane,$^{3,12,13}$
Zouhair Benkhaldoun,$^{14}$
Artem Burdanov,$^{5}$ \newauthor
Francisco J. Pozuelos,$^{3}$
Michael Gillon$^{14}$
Julien de Wit,$^{5}$
Simon F. Green,$^{15}$
Philippe Bendjoya,$^{16}$\newauthor
Jean-Pierre Rivet,$^{16}$ 
Luy Abe,$^{16}$
David Vernet,$^{16}$
Colin Orion Chandler,$^{6}$
Chadwick A. Trujillo$^{6}$
\\ 
$^{1}$Arecibo Observatory, University of Central Florida, HC-3 Box 53995, Arecibo, PR 00612, USA\\
$^{2}$Aix Marseille Universit\'{e}, CNRS, LAM (Laboratoire d'Astrophysique de Marseille) UMR 7326, 13388, Marseille, France \\ 
$^{3}$Space sciences, Technologies \& Astrophysics Research (STAR) Institute University of Li\`{e}ge All\'{e}e du 6 Ao\^{u}t 19, 4000 Li\`{e}ge, Belgium\\
$^{4}$Lowell Observatory, 1400 West Mars Hill Road, Flagstaff, AZ 86001, USA \\ 
$^{5}$Department of Earth, Atmospheric and Planetary Sciences, MIT, 77 Massachusetts Avenue, Cambridge, MA 02139, USA\\
$^{6}$Department of Astronomy and Planetary Science, Northern Arizona University, P.O. Box 6010, Flagstaff, AZ 86011\\
$^{7}$Jet Propulsion Laboratory, California Institute of Technology, 4800 Oak Grove Dr., Pasadena, CA 91109, USA\\
$^{8}$ESA NEO Coordination Centre, Largo Galileo Galilei, 1, 00044 Frascati (RM), Italy\\
$^{9}$Institute for Astronomy, University of Edinburgh, Royal Observatory, Edinburgh EH9 3HJ, UK\\
$^{10}$Armagh Observatory and Planetarium, College Hill, Armagh BT61 9DG, Northern Ireland, UK\\
$^{11}$Institute of Astronomy and National Astronomical Observatory, Bulgarian Academy of Sciences, 72, Tsarigradsko Chauss\`{e}e Blvd., Sofia BG-1784, Bulgaria\\
$^{12}$European Southern Observatory, Alonso de Cordova 3107, Vitacura, Santiago, Chile\\
$^{13}$Oukaimeden Observatory, High Energy Physics and Astrophysics Laboratory, Cadi Ayyad University, Marrakech, Morocco\\
$^{14}$Astrobiology Research Unit, University of Liège, Allée du 6 Août, 19, 4000 Liège, Sart-Tilman, Belgium\\
$^{15}$The Open University, Walton Hall, Milton Keynes MK7 6AA, UK\\
$^{16}$Universit\'{e} C\^{o}te d'Azur, Observatoire de la C\^{o}te d'Azur, CNRS, Laboratoire Lagrange, Campus Valrose, Nice, France\\}

\date{Accepted XXX. Received YYY; in original form ZZZ}

\pubyear{2021}

\begin{document}
\label{firstpage}
\pagerange{\pageref{firstpage}--\pageref{lastpage}}
\maketitle

\begin{abstract}
In December 2018, the main-belt asteroid (6478)~Gault was reported to display activity. Gault is an asteroid belonging to the Phocaea dynamical family and was not previously known to be active, nor was any other member of the Phocaea family. In this work we present the results of photometric and spectroscopic observations that commenced soon after the discovery of activity. We obtained observations over two apparitions to monitor its activity, rotation period, composition, and possible non-gravitational orbital evolution. We find that Gault has a rotation period of $P = 2.4929 \pm 0.0003$ hours with a lightcurve amplitude of $0.06$ magnitude. This short rotation period close to the spin barrier limit is consistent with Gault having a density no smaller than $\rho = 1.85$~g/cm$^3$ and its activity being triggered by the YORP spin-up mechanism. Analysis of the Gault phase curve over phase angles ranging from $0.4^{\circ}$ to $23.6^{\circ}$ provides an absolute magnitude of $H = 14.81 \pm 0.04$, $G1=0.25 \pm 0.07$, and $G2= 0.38 \pm 0.04$. Model fits to the phase curve find the surface regolith grain size constrained between 100-500 $\rm{\mu}$m. Using relations between the phase curve and albedo we determine that the geometrical albedo of Gault is $p_{\rm v} = 0.26 \pm 0.05$ corresponding to an equivalent diameter of $D = 2.8^{+0.4}_{-0.2}$ km. Our spectroscopic observations are all consistent with an ordinary chondrite-like composition (S, or Q-type in the Bus-DeMeo taxonomic classification). A search through archival photographic plate surveys found previously unidentified detections of Gault dating back to 1957 and 1958. Only the latter had been digitized, which we measured to nearly double the observation arc of Gault. Finally, we did not find any signal of activity during the 2020 apparition or non-gravitational effects on its orbit.
\end{abstract}

\begin{keywords}
minor planets, asteroids: individual: (6478) Gault
\end{keywords}



\section{Introduction} \label{sec:intro}

(6478)~Gault (hereinafter Gault) is an asteroid with a previously estimated size of 4 km \citep{Sanchez_2019} located in the main asteroid belt. Dynamically, Gault belongs to the (25)~Phocaea family \citep{Nesvorny_2015}. However this region is composed of two overlapping dynamical families, the (25)~Phocaea dominated by silicaceous S-type asteroids \citep{Carvano_2001} and the (326)~Tamara dominated by carbonaceous C-type asteroids \citep{Novakovic_2017}.

In December 2018, the Asteroid Terrestrial-impact Last Alert System (ATLAS) survey \citep{Tonry_2018} reported that Gault had started showing signs of activity \citep{smith6478Gault2019}. At the time of this discovery, Gault displayed one narrow tail that likely originated on 2018 October $18\pm5$ d according to \citet{Ye_2019} or 2018 October $28\pm5$ d according to \citet{Jewitt_2019}. Additional outbursts of activity were observed around 2018 December $24 \pm  1$ d \citep{Jehin_2019,Ye_2019} or 2018 December $31 \pm  5$ d \citep{Jewitt_2019} and a second tail clearly developed. A last smaller outburst occurred on 2019 February $10 \pm  7$ d \citep{Jewitt_2019}. \citet{Moreno_2019} analysed the activity related to the first two outburst events and found that at least $1.4 \times 10^7$ kg and $1.6 \times 10^6$ kg of dust were released during the respective outbursts. The size of the ejected particles were found to be micrometer to centimeter in size \citep{Moreno_2019,Ye_2019}. Analysis of archival data has further revealed that Gault has displayed episodic activity dating back to at least the year 2013 \citep{Chandler_2019}.

The mechanism that triggered Gault's activity remains unknown, but several hypotheses have been suggested. In the Solar System the main mechanism triggering cometary-like activity is the sublimation of volatiles. However, in the case of Gault, no volatiles and only dust has been detected in the tails \citep{Jewitt_2019}. \citet{Ferrin_2019} hypothesised that Gault could be a comet surrounded by a thick layer of dust that would slowly sublimate as the perihelion distance of Gault decreases with time (due to secular evolution of its orbit). However, activity has been seen throughout Gault's orbit with apparent increased activity near aphelion \citep{Chandler_2019}, suggesting that the activity is not a temperature-dependent, volatile-driven process. 

A second mechanism capable of triggering activity involves impact with a smaller object. 
However, this activity mechanism is incompatible with the multiple episodes of activity observed in the case of Gault as the probability of having multiple impacts in such a short time window would be exceedingly low.

A third mechanism could involve thermal fracturing \citep{Delbo_2014}. While this type of activity may best explain activity for an asteroid with a small perihelion distance like Phaethon \citep{Jewitt_2010}, it is unlikely for Gault. Unlike Gault, Phaethon only shows activity for a few days around perihelion while Gault has been active throughout its orbit. Moreover, Phaethon's perihelion distance is more than 10 times smaller than Gault's, resulting in much greater thermal stress. 

A final and most plausible scenario for activity is that Gault is in a state of rotational instability, perhaps caused by the Yarkovsky–O'Keefe–Radzievskii–Paddack (YORP) \citep{Bottke_2006} spin-up effect. Asteroids are believed to be loosely-bound aggregates (i.e. rubble piles) held together by self-gravity and cohesion between the aggregate particles. This loose agglomeration will disrupt, due to centripetal acceleration, when the rotation rate hits a critical value determined by shape and density \citep{Pravec_2000}. This rotation period is called the spin barrier, as the vast majority of asteroids larger than $\sim$150 m do not rotate faster than a period of $\sim$2.2 hours. The YORP effect is a thermal process that slowly modifies the rotation state (spin and obliquity) of an asteroid. For an asteroid spinning-up under the influence of YORP, it can reach a period close to the spin barrier limit and eject boulders from its surface or even fission into a multi-body system \citep{Jacobson_2011}. This rotational instability was proposed by \citet{Kleyna_2019} to explain Gault's activity, which seemed to be consistent with their finding of a rotation period close to 2 hours. A related scenario to explain Gault's activity would involve a late stage in the process of YORP-driven rotational evolution. Gault may have already suffered rotational disruption and could currently have an undetected bound satellite. As explained in \citet{Jacobson_2011} this satellite could be in an eccentric and chaotic orbit. In such an orbit, during pericenter passages, the secondary could shed mass due to tidal interaction with the primary. 

Before the discovery of Gault's activity in 2018, there were no previously reported lightcurves or spectra. During the 2018-2019 apparition, the majority of attempts to determine the rotational period failed due to one or more of the following possibilities: (1) the presence of dust obscuring Gault's nucleus, (2) a spherical shape that would display light curve variability below detection thresholds, or (3) a rotation axis pointing toward the Earth. Though multiple rotation period estimates have been published \citep{Ferrin_2019,Kleyna_2019,Carbognani_2020}, these results show low amplitude lightcurves dominated by noise, thus making reliable determination of the period  difficult. 
\citet{Kleyna_2019} reported a rotation period around ~2 hours while \citet{Ferrin_2019} reported a rotation period of $3.360 \pm  0.005 $ hours and \citet{Carbognani_2020} a rotation period of $3.34 \pm 0.02$. However, \citet{Ferrin_2019} noted that the lightcurve from \citet{Carbognani_2020} differs from theirs and they interpreted this difference as due to the presence of a secondary. More recently (after submission of the present manuscript) \citet{Purdum_2021} reported a suggestion of a rotation period around 2.5 h while \citet{Luu_2021} is reporting a $2.55\pm 0.1$~h period.  

Uncertainty has also surrounded the interpretation of Gault's composition. Several authors have tried to determine the spectral type of Gault in different taxonomic systems \citep[e.g.][]{Bus_2002,Demeo_2009}. \citet{Jewitt_2019} reported a C-type classification while \citet{Sanchez_2019} reported an S-type. It is interesting to note that these two classes correspond to the taxonomic type expected for members of the Tamara (C-type) and Phocaea (S-type) families. Moreover, \citet{Marsset_2019} and \citet{Lin_2020} both reported a Q-type while the former is observing strong variation of the spectra observed at different epochs. Current space-weathering theories postulate that ordinary chondrite-like asteroids spectroscopically start as Q-types and then gradually turn into S-types due to space-weathering \citep{Mcfadden_1985}.  Q-types are rare in the Main Belt with the exception of the young families \citep{Thomas_2011}. If Gault is compositionally similar to an ordinary chondrite meteorite, we might expect it to be taxonomically closer to Q-type since it is displaying activity that could refresh its surface and suppress space-weathering.  

\section{Observations}

Our first observation of Gault was obtained on 2019 January 8 shortly after it was discovered to be active in late 2018. Since then, we have monitored it over two apparitions through 2021 January 4. The goals of these observations were to monitor the evolution of its activity (see \cite{Moreno_2019} for previously published results), as well as evolution of its spectrum, lightcurve, and orbit (this work). 

Photometric and astrometric observations were obtained from nine facilities located over a wide range of Earth longitudes (23$^{\circ}$E to 111$^{\circ}$W). This wide range allowed us to follow the asteroid over longer intervals than would be achievable with a single observatory, and helped to break a 24 hour observational cadence that could potentially lead to aliases while searching for the rotation period. 

Spectroscopic observations were obtained at the 4.3~m Lowell Discovery Telescope (LDT) located near Happy Jack, Arizona, USA (Minor Planet Center (MPC) observatory code G37), at the 4.1~m Southern Astrophysical Research (SOAR) Telescope located on Cerro Pach\'{o}n, Chile (MPC code I33), and at the 3.56~m New Technology Telescope (NTT) located at La Silla Observatory, Chile (MPC code 809).

A full summary of all facilities used in this work is presented in Table \ref{Tab:Facilities}.

\begin{table*}
\caption{Summary of all facilities used in this work.}
\label{Tab:Facilities}
\begin{tabular}{c c c c c}
\hline
Facility & Techniques & Diameter &MPC code & Filters\\
 &  & (m) & & \\
 \hline
Lowell Discovery Telescope (LDT) & Photometry, Spectroscopy & 4.3 & G37 & VR, Kron-Cousins V and R, Sloan griz \\
Southern Astrophysical Research (SOAR) & Spectroscopy & 4.1 & I33 & \\
New Technology Telescope (NTT) & Photometry, Spectroscopy & 3.58 & 809 & Gunn r  \\ 
Isaac Newton Telescope (INT) & Photometry & 2.5 & 950 & Sloan r \\
2m RCC Rozhen & Photometry & 2.0 & 071 & R \\
Lowell Hall (42 inch) & Photometry & 1.1 & 688 & VR \\
Centre P\'{e}dagogique Plan\`{e}te et Unviers (C2PU) & Photometry & 1.04 & 010 & Sloan r \\
Artemis & Photometry & 1.0  & Z25 & Exo \\
TRAPPIST-North (TN) & Photometry & 0.6 & Z53 & Exo, BVRcIc \\
TRAPPIST-South (TS) & Photometry & 0.6 & I40 & Exo, BVRcIc  \\
\hline
\end{tabular}
\end{table*} 

\subsection{Photometry}
\label{sec:phot}
The photometric observations of Gault were obtained at seven different facilities (Table \ref{Tab:Facilities}). At the 4.3~m Lowell Discovery Telescope (LDT) we made use of the 6144$\times$6160 pixels Large Monolithic Imager (LMI) \citep{Levine_2012,Bida_2014}. Mounted on the LDT, LMI provides a $12.3'\times12.3'$ field of view with a pixel scale of $0.12''$/pixel (unbinned). During our observations we acquired images using a $3\times3$ binning mode providing a plate scale of $0.36''$/pixel. We used VR (bandpass from $0.522 \pm 0.005$ to $0.697 \pm 0.005$ $\rm{\mu}$m), Sloan griz, and Kron-Cousin V and R filters. 

From the La Silla Observatory, we used the 3.58~m New Technology Telescope (NTT). The NTT in imaging mode is equipped with a $2048\times2048$ pixels CCD camera that provides a plate scale of $0.12''$/pixel corresponding to a field of view of $4.1'\times4.1'$. All our observations were performed using the $2\times2$ binning mode ($0.24''$/pixel) and with a Gunn r filter. 

At Isaac Newton Group of Telescopes (ING) we made use of the 2.5~m telescope Isaac Newton Telescope (INT). The INT is equipped with the Wide Field Camera (WFC) that consist of a mosaic of four thinned EEV 2154x4200 pixels CCDs. With each pixel being 13.5$\rm{\mu}$m in size, the plate scale is $0.33''$/pixel allowing an edge-to-edge field of view of $34.2'$. A sloan r filter was use for the INT observations. 

At the Rozhen observatory, we used a 2~m Ritchey-Chr\'{e}tien-Coud\'{e} telescope with an Andor iKon-L DZ936N BEX2-DD equipped with a E2V CCD42-40 $2048\times2048$ pixels CCD providing a plate-scale of $0.5''$/pixel. The field of view is limited to a $10'\times10'$ field by the use of a field mask. All observations were performed with a Johnson-Cousin R filter.      

At Lowell Observatory we used the 1.1~m f/8 Hall telescope. The CCD is an e2v CCD231 $4096\times4096$ pixels CCD with image-scale $1.1''$/pixel when operated in $3\times3$ binning mode, as was done here. The measurements of Gault were made using the same `VR' filter as the one use with the LDT and 5-minute exposures.

At the Calern observatory we made use of the Omicron 1.04~m telescope from the Centre P\'{e}dagogique Plan\`{e}te et Unviers (C2PU). The telescope was operated with a QSI632ws equipped with a KAF-3200ME sensor of 2184x1472 pixels at the prime focus. Such configuration provides a field of view of $15'\times10'$ with a plate scale of $0.425''$/pixel. A sloan r filter was used for the observations at C2PU.

Artemis is a 1~m telescope located at the Teide observatory on the island of Tenerife, Spain. An Andor iKon-L camera of $2048\times2048$ pixels provides a field of view of $12'\times12'$ and a pixel scale of $0.7''$/pixel when used with a $2\times2$ binning. We performed observations both unfiltered and with the Exo filter characterized by a transmission from 0.5~$\rm{\mu}$m to the NIR beyond the infra-red CCD response limit (i.e. blue blocking filter). 

Both TRAPPIST telescopes are 0.6~m robotic Ritchey-Chr\'{e}tien designs operating at f/8 on German Equatorial mounts \citep{Jehin_2011}. At TRAPPIST-North the camera is an Andor IKONL BEX2 DD ($0.60''$/pixel, $20'\times20'$ field of view) and at TRAPPIST-South it is a FLI ProLine 3041-BB ($0.64''$/pixel, $22'\times22'$ field of view). Images were obtained on both telescopes with a binning of $2\times2$ and with the Exo (same as Artemis) and BVRcIc filters.

All photometric data presented in this work were reduced using the PHOTOMETRYPIPELINE \citep{Mommert_2017}. Different filters sets were used at different observatories, but all images were calibrated to r-band magnitudes in the Pan-STARRS catalog and then transformed to Cousins R. We also obtained observations in different filters to measure the photometric colors of Gault so that the calibration and transformation to Cousins R could be validated. All observations meant for color measurements were calibrated in the band they were observed in.

Table \ref{Tab:Phot_Data} from Appendix \ref{ap:table} summarizes all the photometric observations presented in this work.

\subsection{Spectroscopy}
We obtained spectroscopic observation of Gault in the visible wavelength range at the 4.3~m LDT, the 4.1~m Southern Astrophysical Research (SOAR) telescope and the 3.56~m New Technology Telescope (NTT).

At the LDT, we used the DeVeny spectrograph with a plane reflection grating of 150 lines/mm. It is equipped with a $2048\times512$ e2v CCD42-10 with 13.5 $\rm{\mu}$m pixels probing a resolution of 0.43 nm/pixel and covering a spectral range from 0.32 $\rm{\mu}$m to 1 $\mu \rm{m}$. At SOAR the Goodman High Throughput Spectrograph was used with a grating of 400 lines/mm. This configuration offers a dispersion of 0.1 nm per pixel and a wavelength range from 0.5 $\rm \mu$m to 0.9 $\rm \mu$m. At the NTT, we made use of the EFOSC2 low resolution spectrograph with a grism of 100 lines/mm providing a wavelength range from 0.3185 to 1.094 $\rm \mu$m. It is equipped with a $2048\times2048$ CCD camera providing a dispersion of 0.67 nm/pixel.

All the spectroscopic observations from the LDT and SOAR were reduced using the same Python-based spectral reduction pipeline. This pipeline has been fully vetted and been used to reduce hundreds of spectra \citep[e.g.][]{Devogele_2019,Moskovitz_2019,Devogele_2020}. All images are first bias and flat field corrected using a series of 11 individual images to construct the master bias or flat. The flat fields were obtained by uniformly illuminating a white screen located in the dome. The background is corrected by assessing the intensity of the sky on either side of the spectrum for each individual column (spatial dimension). This step is done by fitting a linear slope to the data on both sides of the spectrum and using a 3 sigma clipping procedure to eliminate any pixel that is affected by the spectrum itself or cosmic rays. Then the fit is evaluated for each pixel and subtracted off. This method allows for taking into account spatial variation of the background. Each individual spectrum (asteroid and solar analog) is then extracted, wavelength calibrated, and combined. Ultimately, the last step consists of correcting the asteroid spectrum from the solar component by dividing it by the solar analog spectrum. The solar analog was always observed right before or after Gault and was chosen to match Gault's airmass as closely as possible. To correct the solar component in the measured signal and remove telluric features, the spectrum of the solar analog is gradually shifted (shift of the order of 10$^{-5}$ $\rm \mu$m) with respect to the spectrum of Gault in order to find the dispersion offset that provides the best correction of the telluric absorption features. This process is highly effective at correcting  small relative shifts in the wavelength solutions between Gault and the solar analog. This is because telluric absorption features are very sharp (i.e. have narrow band width), thus a small wavelength shift can introduce strong spikes in divided spectra. 

The spectra taken with the 3.56~m NTT telescope at la Silla were reduced using IRAF procedures for preprocessing, spatial transformation and wavelength calibration, and 1D extraction. The sky background was interpolated from regions located far enough on both sides of the targets.

All the spectroscopic observations are summarized in Table \ref{Tab:Spec_Data}.

 \begin{table*}
\caption{Summary of Gault and solar analog (SA) spectroscopic observations. The V magnitude is the magnitude computed by the Minor Planet center and does not take into account the enhancement due to potential activity}
\label{Tab:Spec_Data}
\begin{tabular}{ccccccc}
\hline
Date & V  & Facility & Int. time   & airmass & SA & Airmass (SA) \\
 &(mag) & & (s) & & & \\
 \hline
2019-01-17 & 19.0 & SOAR & 1200 & 1.05 & SA102-1081 & 1.16 \\
2019-02-15 & 17.7 & SOAR & 3600 & 1.22 & SA102-1081 & 1.71 \\
2019-03-16 & 17.5  & SOAR & 1800 & 1.13 & SA102-1081 & 1.17 \\
2019-03-26 & 17.7 & SOAR & 1800 & 1.18 & SA102-1081 & 1.18 \\
2019-04-06 & 17.9 & NTT & 1800 & 1.37  & SA102-1081 & 1.14 \\
2020-09-11 & 17.2 & LDT  & 1440 & 1.20 & SA93-101  & 1.21 \\
2020-10-05 & 17.1 & LDT  & 1800 & 1.30 & SA93-101  & 1.21 \\
\hline
\end{tabular}
\end{table*}

\subsection{Astrometric and archival observations}
\label{sec:astrom}

Gault has displayed activity not only in 2019, but also in 2016 and 2013 \citep{Chandler_2019}. As these outbursts may have affected Gault's orbit, we investigate its long term orbital evolution by fitting available astrometric data. Non-gravitational mechanisms unrelated to activity can also modify the orbit of an asteroid, for example the Yarkovsky effect \citep{Vokrouhlicky_2000}. To probe the non-gravitational evolution of Gault's orbit we analyzed astrometric data over the longest possible arc (time between the first and last observations). 

This astrometric analysis included a search for previously unidentified observations of Gault on photographic plates and the re-measurement of old observations. The importance of these searches are many-fold. First, we can remeasure the astrometry of old observations using new catalogues such as Gaia DR2 \citep{Gaia_2018}. The use of Gaia DR2 can result in a non-neglectable improvement compared to measurements based on older catalogs. However, the logged time stamps associated with photographic plates can have significant errors. Exposures with photographic plates were often several tens of minutes with time stamps sometimes reported to a precision of the order of a few minutes. For observations dating back many decades, non-specification of the time reference standard can also be an issue. These timing issues can be mitigated by measuring the positions of numbered asteroids whose ephemeris is well enough known that the time of an exposure can be empirically derived with a relatively high precision.

Our astrometric analysis began by remeasuring the first two observations of Gault reported to the MPC. These measurements came from a single 60-minute  exposure from the SERC-EJ survey plate ID 9004 (emulsion IIIaJ plus GG395 filter) taken on 30 January 1984. On this plate Gault is trailed by about $20''$  and the two measurements correspond to the beginning and end of the trail. The fits header on this image reports a time-stamp of 1984-01-30T16:10, which rounded to the closest minute makes the time uncertainty the largest source of error in the astrometric measurements. We used five field asteroids (see Table \ref{Tab:Ast_Meas}) to more accurately determine the mid-exposure time. Our analysis resulted in a mid-exposure time of 1984-01-30T16:09:03 $\pm$ 40~s, which is consistent with the reported time-stamp in the header. Fitting Gault's orbit using the JPL Comet and Asteroid Orbit Determination Package, which is based on standard, state-of-the-art asteroid orbit fitting methods \citep{Farnocchia_2015},
the new measurements possesses a residual (differences between the measurement and the best fitted orbit) of $0.33''$ while the two old measurements had residuals of $0.84''$ and $1.51''$. The new measurements for Gault and the other asteroids are reported in Table~\ref{Tab:Ast_Meas}. 

We also performed a plate search for new detections of Gault by cross matching its ephemeris against catalogs from about half a dozen photographic plate surveys. Amongst those surveys was the first Palomar Observatory Sky Survey (POSS-I). POSS-I operated for about 10 years from the end of 1949 through the end of 1958. This survey obtained pairs of red and blue-optimized plates across a field of view of $6^{\circ} \times 6^{\circ}$ with sky coverage from $+90^{\circ}$ to $-27^{\circ}$ declination. This archival search yielded a detection of Gault on the POSS-I blue plate (103aO emulsion), plate number 1619 (8 minute exposure) obtained on 1958 December 11. Ambiguities about the exact time of acquisition, possibly due to transcription errors in the time stamps for the red and blue 1619 plates, were resolved by again measuring the positions of numbered field asteroids with well-defined ephemerides (Table \ref{Tab:Ast_Meas}). We thus obtained a precise mid-exposure time of 1958-12-11T09:53:30$\pm$16~s. This observation increased the length of Gault's orbital arc by a factor of 1.7, allowing for significantly improved accuracy in the orbit determination and a search for non-gravitational effects. 

Our search also identified a visual detection of Gault on plate number 3055 taken with the camera on the 13-inch Pluto discovery telescope at Lowell Observatory. Gault was near the detection threshold of the plate, predicted to be around B$\sim18.3$. The date of this observation was 1957 September 3, making it the oldest known detection of Gault. However, this plate is not digitized. As such retrieval of astrometry is beyond the scope of this work and will be deferred until the Pluto camera plates are digitized.

\begin{table}
\caption{Archival data astrometric measurements}
\label{Tab:Ast_Meas}
\begin{tabular}{c c c c }
\hline
Asteroid & RA  & Dec & Uncertainty \\
 & (hms)  & (dms) & ($''$)\\
 \hline
\multicolumn{4}{c}{SERC-EJ plate ID 9004: 1984-01-30T16:09:03$\pm$40 s}  \\ 
\hline
(6478)~Gault  & 10 50 48.136  & -12 22 12.97 & 0.20\\
(25731) 2000 AL193 & 10 49 12.295 & -12 19 45.37 & 0.20\\
(40902) 1999 TY143 & 10 52 55.651 & -13 07 23.43 & 0.37 \\
(68923) 2002 LV40  & 10 47 06.124  & -12 25 18.65 & 0.20\\
(130243) 2000 CA75 & 10 51 34.376 & -12 08 03.97 & 0.12\\
(161547) 2004 XS29 & 10 49 27.892 & -13 33 17.60 & 0.22\\
\hline
\multicolumn{4}{c}{POSS I 103aO plate ID 1619: 1958-12-11T09:53:30$\pm$16 s } \\ 
\hline
(6478)~Gault & 07 28 59.526 & -10 50 12.71 & 0.20\\
(26303)~1998 SD144 & 07 29 37.216 & -11 38 32.70 & 0.14\\
(29945)~1999 JU83  & 07 22 07.370 & -10 54 53.16 & 0.14\\
\hline
\end{tabular}
\end{table} 

\begin{figure}
\centering
\includegraphics[width=9cm]{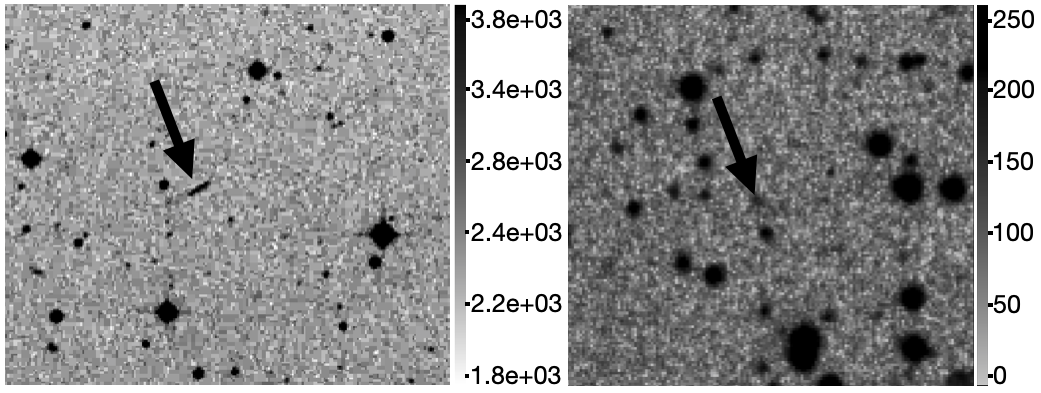}
\caption{Archived observations of Gault. The left figure shows the SERC-J survey plate ID 9004 obtained on 30 January 1984. The acquisition time was 60 minutes and Gault was moving fast enough to appear as a continuous trail. The right figure shows the oldest  measured observation of Gault from the POSS-I plate ID1619 blue plate obtained on 11 December 1958. This measurement increased Gault's arc by a factor of 1.7.}
\label{fig:Gault_Period}
\end{figure}

\section{Results and discussion}

\subsection{Rotation period}
\label{Subsec.Period}

During the 2019 apparition, the nucleus was obscured behind  dust, thus making detection of the rotational lightcurve difficult.
As such multiple suggestions of Gault's rotation period have appeared in the literature, ranging from 2 to 3.6 hours \citep{Kleyna_2019,Carbognani_2020,Ferrin_2019,Purdum_2021,Luu_2021}. The analysis of our data obtained in 2019 did not show any significant periodic signal. 

On the other hand, Gault did not show any detectable sign of activity during the 2020 apparition (\S\ref{Sec:Activity}). We were thus able to obtain very high signal-to-noise data throughout the apparition. 

In assessing the rotation period of Gault we use photometry from the LDT, Rozhen, TRAPPIST-North from 2020 August 22, and the Lowell 42-inch from 2020 September 15. We limited this analysis to just these lightcurves as they provide the largest temporal coverage and the highest signal to noise. Even though the TRAPPIST data were from the smallest telescope used in this study, they are extending the arc of observations and improve the period determination. With the first lightcurve obtained on 2020 August 22 and the last one on 2020 November 18, the lightcurves span 87 days. Gault still displays a low amplitude lightcurve during 2020 apparition and to improve the period determination, all observations with a computed uncertainty larger than 0.02 mag were discarded from this analysis. To account for the varying magnitude due to changing viewing geometry, we normalized all individual lightcurves to their respective means.

\begin{figure*}
\centering
\includegraphics[width=18cm]{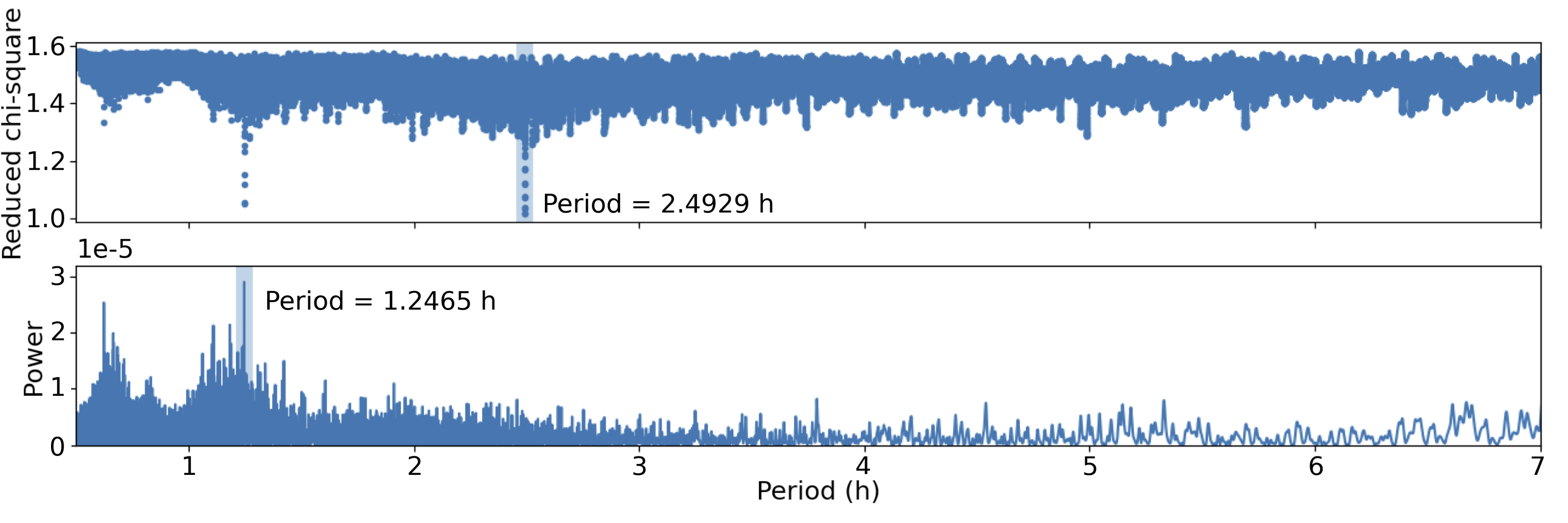}
\caption{Top panel: reduced chi-square residuals from the fits of a fourth-order Fourier series to the lightcurves. The best fit corresponds to a rotational period of $2.4929 \pm 0.0003$ h.
Lower panel: Lomb–Scargle periodogram. The peak in the periodogram is at $1.2465 \pm 0.0003$ h, corresponding to half the value of the best fit period. The relatively high reduced chi-squared is due to the complicated shape of the Gault's lightcurve and its variation over time.}
\label{fig:Gault_LS_2020}
\end{figure*}

We performed two different analyses to find the rotation period. A Lomb-Scargle periodogram \citep{Lomb_1976,Scargle_1982} (bottom panel of Figure \ref{fig:Gault_LS_2020}) shows a strong peak around P = $1.2465 \pm 0.0003 $ h and a secondary peak at P = $0.62318 \pm 0.00009$ h. 
However, due to the dual peaked morphology seen for the majority of asteroid lightcurves, the Lomb-Scargle periodogram most often shows a peak at one half the actual rotation period. We thus performed a second analysis by fitting a Fourier series of fourth order to the data while scanning through a wide range of fixed periods from $0.024$~h to 7.2~h in step sizes of $3\times 10^{-6}$~h (top panel of Figure \ref{fig:Gault_LS_2020}). At each period step we computed a reduced chi-squared statistic. This Fourier analysis clearly shows a minimum in the reduced chi-square for a period of $2.4929 \pm 0.0003$~h corresponding to twice the 1.2465~h and four times the 0.62318~h periods from the Lomb-Scargle analysis. 

\begin{figure*}
\centering
\includegraphics[width=18cm]{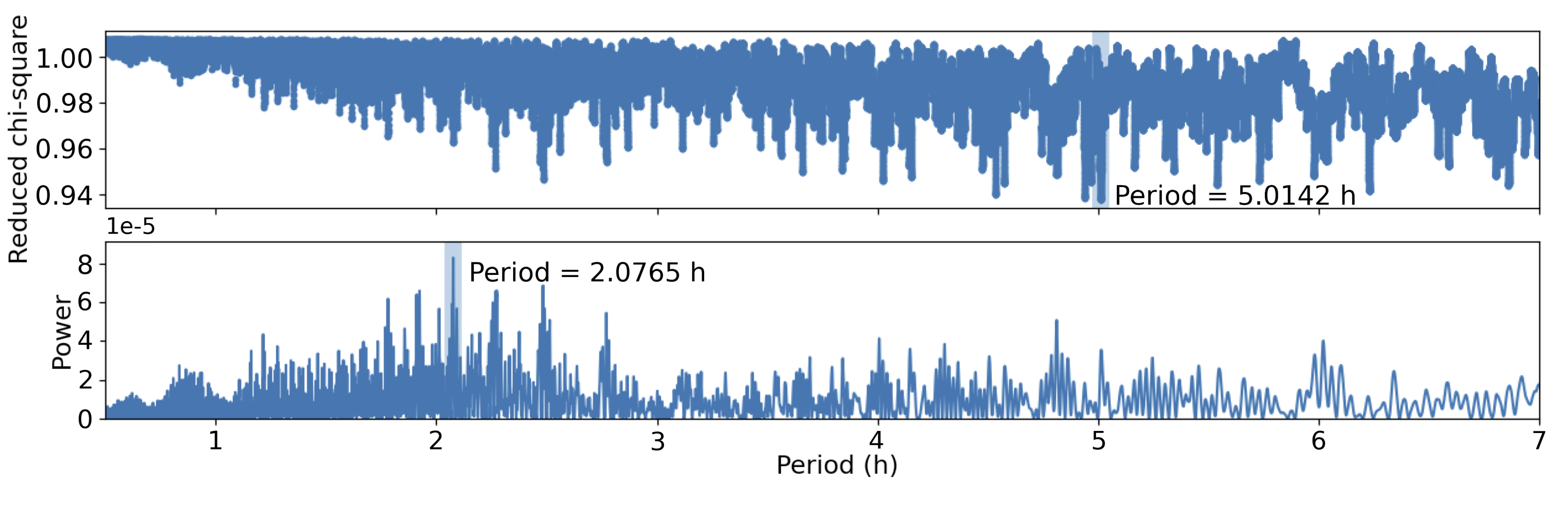}
\caption{Same as Figure \ref{fig:Gault_LS_2020}, but for the 2019 apparition, no clear period can be derived from this analysis.}
\label{fig:Gault_LS_2019}
\end{figure*}

We performed the same analysis but considering just data from the 2019 apparition. The highest peak in the Lomb-Scargle periodogram corresponds to a period of 2.0765 hours. However, we found that using different uncertainty thresholds for the exclusion of data yielded different answer for the highest peak. This suggests that the 2019 apparition cannot be solely used to determine the rotation period of Gault. Moreover, the Fourier analysis provides a rotation period of 5.0142~h incompatible with the one obtained with the Lomb-Scargle periodogram. As explained earlier, we interpret this non-detection of the rotation period of Gault as a result of photometric contamination by the tail. Both periodograms are displayed in Figure \ref{fig:Gault_LS_2019}.

\begin{figure*}
\centering
\includegraphics[width=18cm]{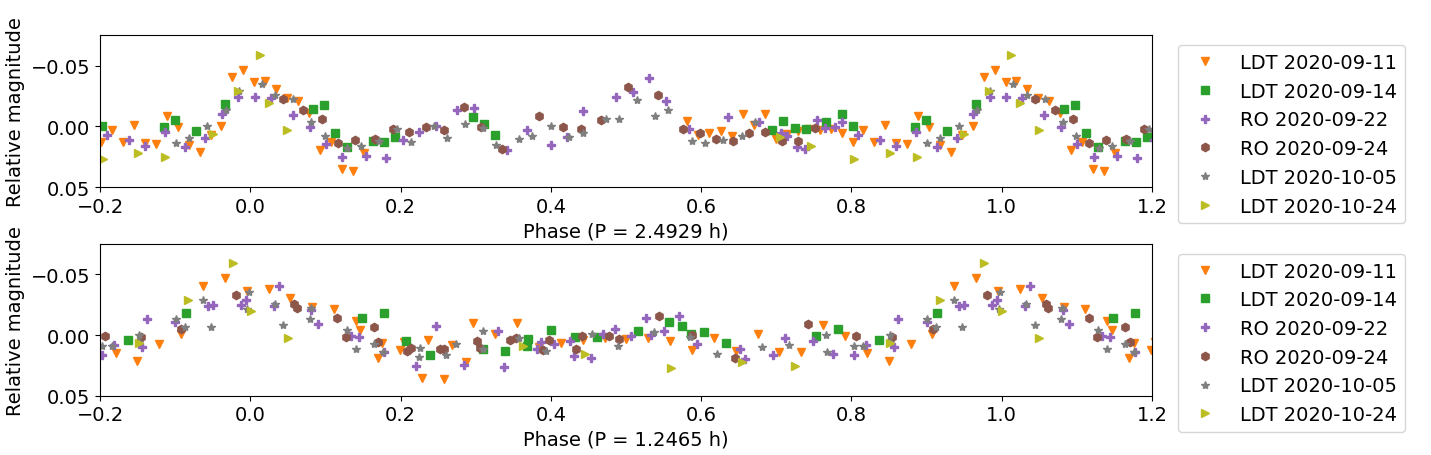}
\caption{Upper panel: rotational lightcurve of Gault folded to a period of 2.4929 h. Lower panel: rotational lightcurve of Gault folded to a period of 1.2465~h.}
\label{fig:Gault_Period_2020}
\end{figure*}

\begin{figure}
\centering
\includegraphics[width=9cm]{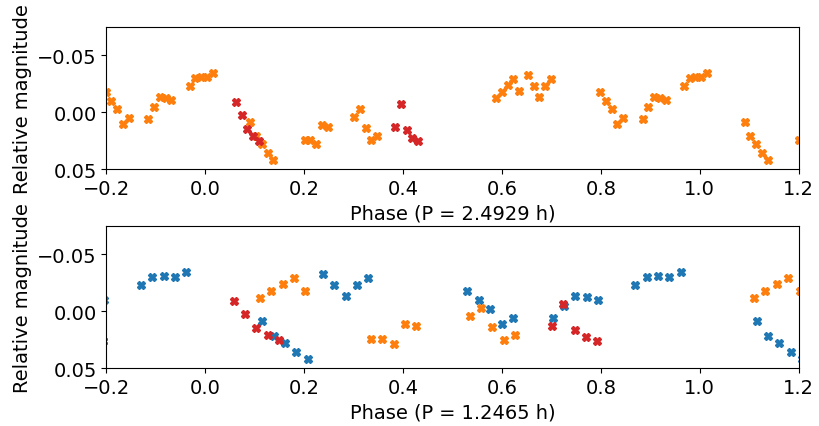}
\caption{Last observation of Gault obtained with the LDT on 2020 November 18. The upper panel shows them folded with a period of 2.4929~h while the lower panel shows them folded with a period of 1.2465~h. Since the observation spans more than 3h, different cycles are represented in different colors. The preference for the 2.4929~h is made clear in this case.}
\label{fig:Gault_Period_Last}
\end{figure}

Figure \ref{fig:Gault_Period_2020} shows our photometry for the 2020 apparition folded to the 2.4929~h (top-panel) and to the 1.2465~h (bottom) periods. Based on the folded lightcurves, both appear to be reasonable solutions. However Figure \ref{fig:Gault_Period_Last} represents an equivalent to Figure \ref{fig:Gault_Period_2020}, but for the LDT 2020 November 18 lightcurve only. These data were excluded from Figure \ref{fig:Gault_Period_2020} due to the changed morphology of the lightcuve on this date, which was likely due to the higher solar phase angle of $20.9^{\circ}$ associated with these particular observations. Folding these data to the 1.2465~h period clearly demonstrates inconsistent trends in the measured magnitudes, for example around rotational phases of 0.3 and 0.9. The 2020 November 18 thus provides strong evidence for distinguishing between the two periods.

Another argument for rejecting the 1.2465~h period comes from the fact that it would be in contradiction with the observed spin-barrier at $\sim P = 2.2$~h for asteroids larger than $\sim 150$~m \citep{Pravec_2000}. In order to rotate with a period of 1.2465~h, Gault would have to possess significant internal strength (see \S\ref{sssec:YORP}).

\subsection{Taxonomy, Color and spectral slope}
\label{sec:color}
We obtained spectral observation of Gault during two separate apparitions, both when it was displaying activity and not. We also obtained photometric observations to determine its broad-band color.

\begin{figure*}
\centering
\includegraphics[width=18cm]{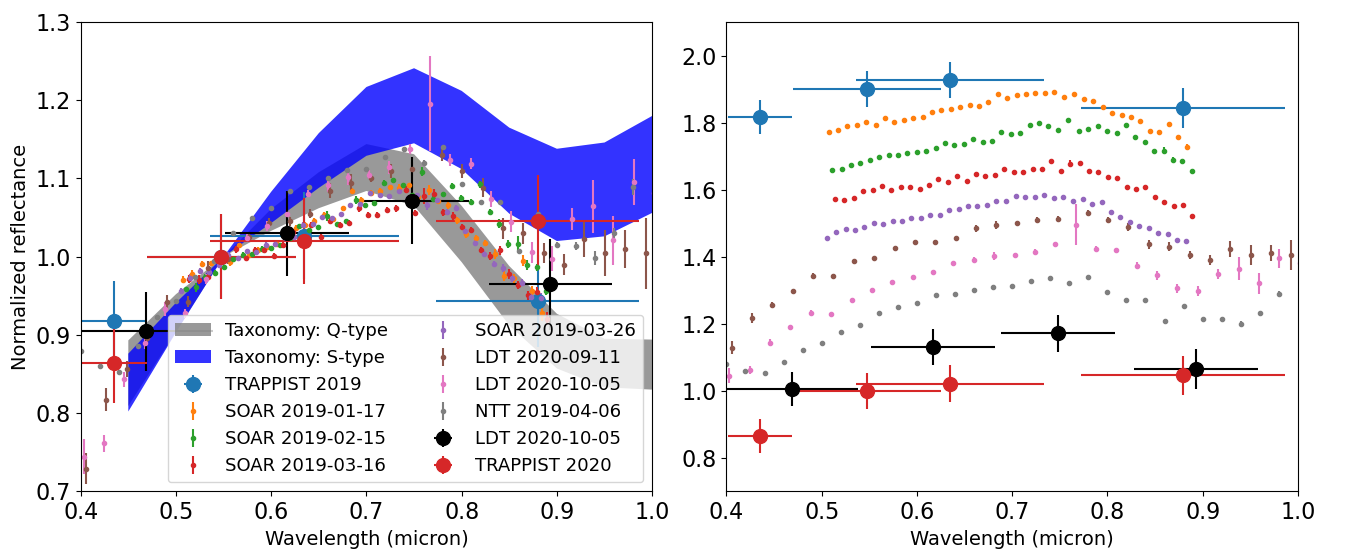}
\caption{Left: All visible spectra and photometric colors presented in this work compared with \citet{Demeo_2009} spectral taxonomic envelopes of the S and Q-type. Right: Same as the left part, but individual spectra have been shifted in reflectance for visualization. Gault spectra is intermediate between S and Q-type, but the spectra showing the longest wavelength range to the near-infrared are indicative of a S-type asteroids. All color and symbols references are the same as in the left part of the figure.}

\label{fig:All_Spec}
\end{figure*} 

Figure \ref{fig:All_Spec} displays all visible spectra and broad-band colors obtained in this work. The left part of Figure \ref{fig:All_Spec} compares all the spectra with the spectral envelope of the S and Q-types from the \citet{Demeo_2009} taxonomic system. All the Gault spectra are consistent with either one of these classes. The maximum of reflectance for Q-types occurs around 0.7 $\rm{\mu}$m while for S-types this maximum is shifted to longer wavelengths around 0.75 $\rm{\mu}$m. In the case of Gault, the maximum occurs around 0.76 $\rm{\mu}$m, suggesting a low slope S-type rather than a Q-type.
Moreover, the NTT and LDT spectra cover a larger range of wavelength compared to the SOAR spectra. All the LDT and the NTT spectra perfectly match each others and show a shallow 1~$\rm{\mu}$m absorption band characteristic of an S-type rather than a Q-type.

We notice that the four spectra obtained at SOAR during the 2019 apparition all agree within the uncertainties. However they do not agree with the NTT spectra that was obtained a few weeks after the last SOAR spectrum. On the other hand, the NTT agree perfectly with all the 2020 spectra obtained at LDT. Even if the SOAR spectra does not match the other spectra, the differences are only slight variation in slope than can easily be attributed to any observation/calibration issues. However, it is interesting to note that the SOAR spectra have been obtained when Gault was the most active and that Gault did not show any sign of activity when the LDT spectra were obtained in 2020. Unless such small variation in slope as a function of activity can be confirmed when Gault will be active again, we conclude that we are not seeing any variation of the spectral properties as a function of time for Gault.

Due to Gault's recent activity one might expect a fresh Q-type like surface rather than a space-weathered S-type. The fact that our spectra are most consistent with a weathered S-type surface could mean that Gault's activity is localized to a small region on the surface. YORP driven activity would be consistent with activity localized in the equator region while the pole regions could possess older S-type like surfaces.

To investigate the effect of the selected solar analog on our spectra, we observed three different solar analogs during the 2020 November 5 night at LDT. 
Our analysis shows that, even though slope variations in the final spectra may be larger than formal uncertainties, the selection of different solar analogs can not explain the level of variation observed between the SOAR and the LDT and NTT spectra. The small variations observed with different solar analogs can be attributed to differences in airmass or atmospheric conditions. 

\begin{figure}
\centering
\includegraphics[width=9cm]{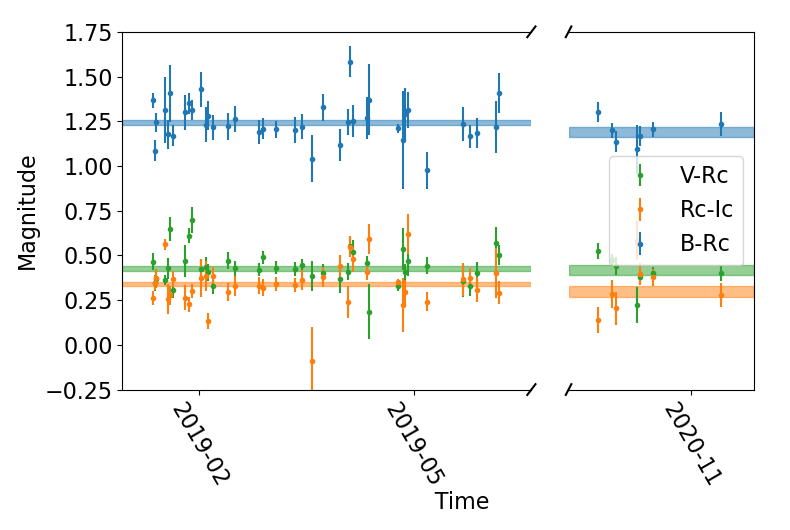}
\caption{V-Rc, Rc-Ic, and B-Rc color indexes of Gault as a function of time. No systematic variation of the colors of Gault is observed. The shaded areas correspond to the estimated uncertainties on the median values of the estimated color for each opposition independently.}
\label{fig:All_Colors}
\end{figure}

During the 2019 and 2020 apparitions we also regularly observed Gault using the B, V, Rc, and Ic filters at both the TRAPPIST-North and South telescopes. Figure \ref{fig:All_Colors} shows the colors indexes V-Rc, Rc-Ic, and B-Rc as a function of time over the two apparitions. No significant variation of the color over time can be observed. We find a mean B-Rc = $1.242 \pm 0.014$, V-Rc = 0.429 $\pm$ 0.013 mag and Rc-Ic = 0.338 $\pm$ 0.012 for the 2019 apparition and B-Rc = $1.189 \pm 0.030$, V-Rc = 0.419 $\pm$ 0.029, Rc-Ic = 0.300 $\pm$ 0.032 for the 2020 apparition. These values are consistent with no color variation from one opposition to the next.

Color observations were also obtained at the LDT during the night of the 2020 November 5. We alternated filters in the following sequence VR-R-VR-V-VR-g-VR-r-VR-i-VR-z. This sequence allowed us to correct observations obtained in R,V,g,r,i, and z filters for lightcurve variations by using the interleaved VR images as a reference to monitor rotational variability. The Kron-Cousins R and V filter were used to obtain a precise V-R = 0.418 $\pm$ 0.014 magnitude. The V-R color found from the LDT is consistent with the TRAPPIST value.    

The V-R color index of Gault is compatible with Q (V-R = 0.424), V (V-R = 0.413), or X (V-R = 0.410) taxonomic types according to \citet{Dandy_2003}. However, it is not compatible with a C or S-type classification that would require a V-R of 0.376 and 0.475 respectively. As discussed earlier, Gault spectra resemble the one of the S-type with a lower than usual spectral slope, translating in a lower V-R colors. The V-R color index is used below to transform our $H$ magnitude determination from R to V band. 

The g,r,i, and z Sloan and the B, V, Rc, and Ic Jhonson-Cousin filters were used to obtain a coarse spectrum of Gault at the LDT and TRAPPIST respectively. The black dots in Figure \ref{fig:All_Spec} represent the LDT spectro-photometry data while the blue and red dots represent the TRAPPIST data for 2019 and 2020 respectively. The horizontal bars represents the wavelength passband of each filter while the vertical bars represent the uncertainty associated with each measurement.

\subsection{Phase curve}
\label{sec:PhaseCurve}
Our continuous monitoring of Gault during both the 2019 and 2020 apparitions allow us to analyse the variation of Gault's magnitude as a function of the solar phase $\alpha$ (i.e. the angle between the Sun, Gault, and the observer) and time. During the 2019 apparition, we observed Gault from $-19.8^{\circ}$ (negative sign for phase angles prior to opposition) to $27.8^{\circ}$ (positive sign for phase angles post opposition) with a minimum phase angle of $6.9^{\circ}$ while during the 2020 apparition, we observed Gault from $-18.4^{\circ}$ to $23.6^{\circ}$ with a minimum at $0.4^{\circ}$. 

The 2019 apparition is characterized by a magnitude variation mainly driven by Gault's activity while the 2020 apparition is characterized by a magnitude variation due to phase angle variation. As already explained in Sec. \S\ref{sec:phot} all our photometric measurements are calibrated in the R band using the Pan-STARRS photometric catalog. This calibration generally involved transforms from the Pan-STARRS ugriz filter set to Johnson-Cousins UBVRI \citep{Chonis_2008}.

\begin{figure}
\centering
\includegraphics[width=9cm]{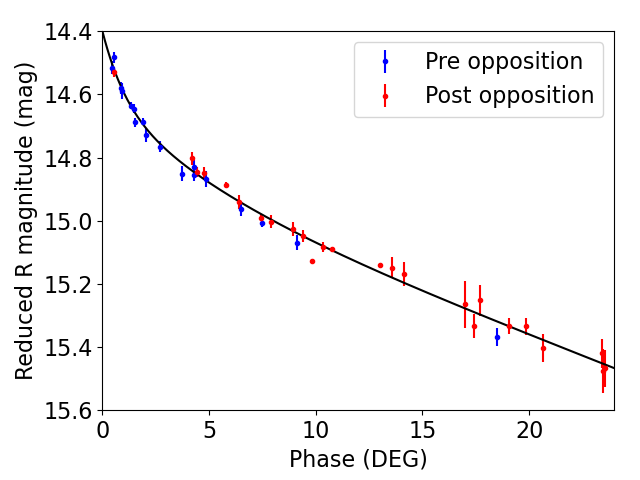}
\caption{Phase curve of Gault calibrated in the R band using the Pan-STARRS photometric catalog. All magnitudes are corrected from the distance of the Gault from the Earth and the Sun. 
The blue and red points respectively represents the pre- and post- opposition measurements.
The pre- and post- opposition measurements are fitted using a single relation showing that Gault's activity was nonexistent or below our detection limits. 
}

\label{fig:Phase_Curve}
\end{figure}

Fig. \ref{fig:Phase_Curve} represents the 2020 apparition reduced magnitude (corrected for distance from the Earth and the Sun) as a function of phase angle. Each point in the phase curve represents the mean magnitude of all observations obtained during a given night. We did not apply any corrections for lightcurve amplitude as it was found to be quite low and because of the short rotation period compared to the typical length of an observing session such that both lightcurve maxima and minima were sampled (\S\ref{Subsec.Period}). The error bars in the phase curve represents the standard deviation of all measurements considered for computation of the mean. We did not divide the value for the error bars by the square root of the number of observations in order to account for the real dispersion due to light-curve variation.

During our observation in 2020, there was no sign of activity and we were able to observed Gault at a large range of phase angles allowing us to model its phase curve using the $H$, $G1$, $G2$ system \citep{Muinonen_2010}. We used a \textit{Python} package from the University of Helsinki to fit our measurements of Gault\footnote{https://wiki.helsinki.fi/display/PSR/HG1G2+tools} \citep{Penttila_2016}. Fig. \ref{fig:Phase_Curve} shows data before and after opposition respectively in blue and red dots so as our phase-curve analysis fits the data prior or after opposition equally well.

The best fit phase curve provides $H_{\rm R} = 14.397 \pm 0.035$ mag, $G1 = 0.25 \pm 0.07$, and $G2 = 0.38 \pm 0.04$. Applying the V-R color index found in the previous section, we find that $H_{\rm V} = 14.81 \pm 0.04$ (the $H$ magnitude is defined in the V band by the IAU). This value is 0.5 magnitude larger than that reported by the JPL Horizons service. Such discrepancies on $H$ are not uncommon, but are often due to lightcurve variation, which is not the case for Gault (\S\ref{Subsec.Period}). We suggest this difference may be attributable to Gault's activity such that some of the magnitudes reported to the MPC are biased toward lower values. A 0.5 difference in H magnitude for an object with a typical S-type albedo results in a decrease of its size by approximately 20\%. 

\cite{Shevchenko_2016} analysed the phase curves in the $H$, $G1$, $G2$ system of 93 asteroids and found a correlation between the $G1$ and $G2$ parameters. We find that our determination of $G1$ and $G2$ for Gault falls on this expected correlation. These authors also found that different taxonomic types fall on different regions of the $G1$-$G2$ parameter space. We find that our values of $G1$ and $G2$ correspond with the one found for S-complex asteroids (there is only one Q-type in \citet{Shevchenko_2016} dataset, but its G1, G2 values are similar to the S-type asteroid values). Namely, S-complex asteroids are characterized by $G1 = 0.26 \pm 0.01$ and $G2 = 0.38 \pm 0.01$ while C-class asteroids are characterized by $G1 = 0.82 \pm 0.02$ and $G2 = 0.02^{+0.02}_{-0.01}$ \citep{Shevchenko_2016}. \cite{Shevchenko_2016} also found a correlation between the $G1$ and $G2$ parameters with albedo. If we apply their calibration we find that Gault has an albedo of $p_{\rm V} = 0.26 \pm 0.05$. Applying the relation existing between the absolute $H$ magnitude and albedo, we find that Gault possesses a diameter of $D=2.8^{+0.4}_{-0.2}$ km

Our observation of Gault cover both pre- (up to $\alpha = 18.5^{\circ}$) and post- (up to $\alpha = 23.6^{\circ}$) opposition. Comparing these pre- and post-opposition data does not reveal any meaningful differences in the phase curve behavior. While the pre-opposition points do appear slightly lower than the post-opposition measurements, this difference (if real) is within expected variations due to changing observing geometry throughout the apparition. Despite the episodic nature of Gault's activity, this result suggests that Gault did not experience any activity during our observations. To further check for activity, we analysed observations reported to the MPC by a few reliable sky surveys and we performed a deep stack of the LDT's observations, our deepest set of imagery(\S \ref{Sec:Activity}).   

\begin{figure}
\centering
\includegraphics[width=9cm]{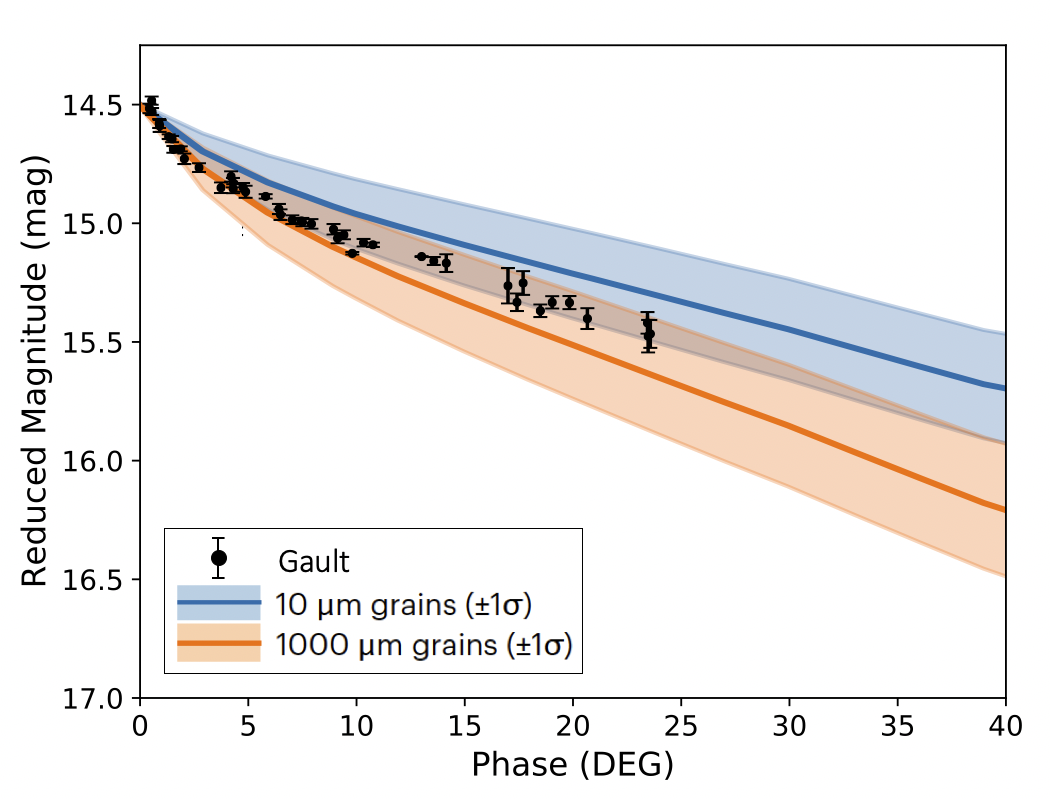}
\caption{Phase curve of Gault calibrated in R (black) plotted against Hapke-derived photometric phase curves for a spherical S-type asteroid at two different surface grain sizes. The shaded regions represent the model variance across the Hapke free parameters (i.e., composition, surface roughness, width strength of opposition surge) used in the model inputs.}
\label{fig:Gault_Hapke_PhaseCurve}
\end{figure}

We also analyzed Gault's phase curve in relation to a suite of model-derived phase curves to constrain the regolith grain size on Gault's surface. We have implemented a Hapke radiative transfer model \citep{Hapke_2012} which describes the bidirectional reflectance properties of regolith surfaces of Solar System bodies \cite[e.g.][]{Grundy_2009, Grundy_2018}. Using this model, we derive Hapke disc-integrated photometric phase curves in R band for a phase angle range of $0^{\circ}$-$40^{\circ}$. Figure~\ref{fig:Gault_Hapke_PhaseCurve} shows the range of Hapke photometric phase curves derived for a spherical S-type asteroid with surface regolith grains of 10 $\mu{\rm m}$ (blue) and 1000 $\mu{\rm m}$ (orange). The shaded regions represent the model variance derived using ordinary chondrite compositions from H to LL \citep{Dunn_2010}, which are typical of S-type asteroids, whose optical constants are derived using the relationship outlined in \cite{2013Trang}. We use a range of surface roughness values from 20$^{\circ}$-40$^{\circ}$, width of the opposition surge from 0.04-0.08, and opposition strength from 1.0-3.0, all representative of S-type asteroids \citep{2015Li} as inputs into the Hapke model. 

The suite of Hapke model results suggest a best fit surface grain size in the range of 100-500 $\mu{\rm m}$ for Gault. These results are consistent with results presented in \cite{Jewitt_2019} where they estimate the mean particle radius in the tail during activity in 2019 to be $\sim$100 $\mu{\rm m}$ while the prediction for particles much closer to the nucleus is $\sim$500 $\mu{\rm m}$ suggested by radiation pressure acceleration. \citet{Moreno_2019} predict the grain size of the dust particles composing the tail are in the range of 1 $\mu{\rm m}$ to 1 mm, also consistent with our predictions from the phase curve fits. 

\subsection{Gault activity}
\label{Sec:Activity}

\subsubsection{Is Gault still active?}

Gault was active in late 2018 through mid 2019, but it remains unclear whether activity was present in 2020. To check for activity, we performed deep stacks of images obtained with the LDT and TRAPPIST. We also analysed the secular variation of Gault's magnitude using our obsvervations, but also observations submitted to the Minor Planet Center. 

\begin{figure*}
\centering
\includegraphics[width=17cm]{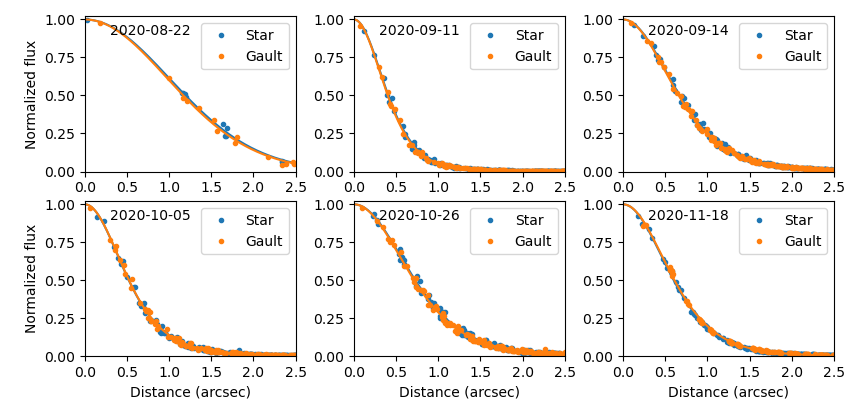}
\caption{Radial profile of Gault (orange) and a comparison star (blue) for five different LDT nights. There is no differences between the PSFs showing that Gault is a point-source like object. The continuous lines corresponds to the radial profile of a 2D MOFFAT function fitted to the PSF of each object. The lines are so similar that the blue fit (comparison star) is hidden behind the orange.}
\label{fig:Gault_RadProf}
\end{figure*}

Our deep stack analysis involved two different stacks. The first one is obtained by stacking on the asteroid, i.e. shifting all images according to the asteroid motion while the second one is performed by stacking on field stars. Our observations were obtained tracking at the asteroid rate, thus, in principle no registration of the images are necessary to create the asteroid-stacked image. However, to account for slight telescope drift, tracking error over the course of the night, and field distortions, we first astrometrically calibrated our fields referencing the Gaia catalog and using the \textit{photometrypipeline} \citep{Mommert_2017}. 
The asteroid motion was retrieved using JPL Horizons ephemeris (updated to take into account our new measurements from 1958 and 1984). To avoid star trailing, the asteroid motion during the length of the exposures was always kept smaller than the mean seeing of the night (from 1.9 to 4.2 time smaller).  The resulting stacks yielded nearly identical PSFs for the asteroid and the stars (see Fig. \ref{fig:Gault_RadProf}).

To compare the PSF profiles of both field stars and Gault, we first measure the centroid of the asteroid and the comparison star by fitting a 2D Moffat function \citep{Moffat_1969} to the profiles using the \textit{astropy.modeling.functional\_models.Moffat2D} function. We then computed the distance of each pixel from the photocenter. To compare the radial profile of the star and the asteroid, we normalized each pixel based on the amplitude of the 2D Moffat. We then plot the normalized flux as a function of the distance to the photocenter. Figure \ref{fig:Gault_RadProf} shows the radial profile of Gault and a field star for the first TRAPPIST night (even if obtained with a small telescope, we here considered the TRAPPIST night of 2020 August 22 as it is the earliest observation in the 2020 apparition) and five LDT nights. For all these nights the radial profile of Gault and comparison star are found to be identical. This shows that Gault is a point-source like object at our level of detection.

With this radial profile analysis we can assess the activity level that could have been observed if present. This is done by photometrically calibrating the image stacked on the field stars. This process is identical to measuring the magnitude of Gault in our light-curve analyses and provides the means to obtain the zero-point magnitude on the stacked image. Since the asteroid stack was obtained using the same images as the star stack, we can assume that the zero-point magnitude is identical for both. We can thus assess the magnitude of the sky-background (zero-point magnitude) and the noise associated with it to derive the faintest coma that would have been detected at a 3 sigma level. We also determined the minimal dust production rate that could have been detected on our images. For this, we followed the procedure detailed in \citet{Jewitt_2013}. Table \ref{Tab:Coma_anal} shows the result of our analysis for the six nights. The deepest images were obtained on 2020 November 18 and show that a R=30.47 mag/$''^2$ coma or tail would have been detected with 3 sigma confidence. On this night the seeing was $1.26''$ which corresponds to a physical size of 1,586 km at Gault's distance. These images provide the deepest sampling due to the fact that the Moon was not up. On the other hand, the 2020 September 11 observations have the best seeing ($0.8''$) which would have allowed to detect coma activity much closer to the asteroid (739 km). If we assume that we need to be at at least 3 seeing disks away from the photocenter to detect a coma with our sky-background limit analysis, then this corresponds to physical extents of 2,217 km and 4,759 km on the nights of September 11 and November 18 respectively. In no cases, we detect any sign of activity and the mimimal production rate that could have been detected is 0.002 kg/s. 

\begin{table*}
\caption{Summary parameters of the activity search}
\label{Tab:Coma_anal}
\begin{tabular}{cccccccccccc}
\hline
Date & R$^{a}$  & $t_{\rm exp} ^{b}$ & Motion$^{c}$ & \#$^{d}$   & seeing$^{e}$ & $B_{\rm RMS}^{f}$ & ZP$^{g}$ & $m_{\rm{surf}}^{h}$  & Dist$^{i}$  & $M_{\rm{r}}^{j}$ & Telescope \\
 & (ADU)  & (s) & ($''$) &  &($''$) &(mag) &(mag) & (mag/$''^{2}$) & (km) &(kg/s) & \\
\hline
2020-08-22 & 17.74 & 120 & 1.0 & 94 & 2.36 & 6.05 & 28.60 & 25.06 & 6963 & 0.48& TN\\
2020-09-11 & 17.35 & 30 & 0.42 & 120 & 0.80 & 2.45 & 28.40 & 28.46 & 2217 & 0.005 & LDT \\
2020-09-14 & 17.29 & 30 & 0.44 & 65  & 1.31 & 1.78 & 29.19 & 29.59 & 3620 & 0.003 &  LDT\\
2020-10-05 & 17.24 & 20 & 0.28 & 105 & 1.05 & 3.50 & 28.81 & 28.48 & 2981 & 0.005 & LDT \\
2020-10-24 & 17.70 & 60 & 0.54 & 30  & 1.43 & 3.63 & 28.92 & 28.92 & 4492 & 0.007 & LDT\\
2020-11-18 & 18.47 & 60 & 0.30 & 50  & 1.26 & 2.98 & 30.63 & 30.47 & 4759  & 0.002 & LDT\\
\hline
\end{tabular}
\\
      \small
    $^{a}$ Measured R magnitude, $^{b}$ Exposure time of individual frames, $^{c}$ Motion on the sky in arcsecond of Gault during one acquisition, $^{d}$ Number of frame in the stack, $^{e}$ Average measured FWHM of Gault and the comparison star on the stacks, $^{f}$ Standard deviation of the sky background surrounding Gault and the comparison star on the stacks, $^{g}$ Zero point measured on the star stack, $^{h}$ Surface magnitude of the sky background around Gault and the comparision star on the stacks, $^{i}$ Minimum distance from the nucleus from which coma detection would be possible at the given precision (3 times the PSF's FWHM projected at Gault distance from Earth), $^{j}$ Minimal mass loss rate in dust in kg/s that would have been detectable following the method of \citet{Jewitt_2013}
\end{table*} 

The presence of activity can also be detected by monitoring the evolution of Gault's magnitude over time. To monitor the magnitude variation due to activity, we first corrected the magnitude from the distance of Gault from the Earth and the Sun and then corrected from the phase angle variation using the phase curve model derived in \S\ref{sec:PhaseCurve}.

\begin{figure*}
\centering
\includegraphics[width=17cm]{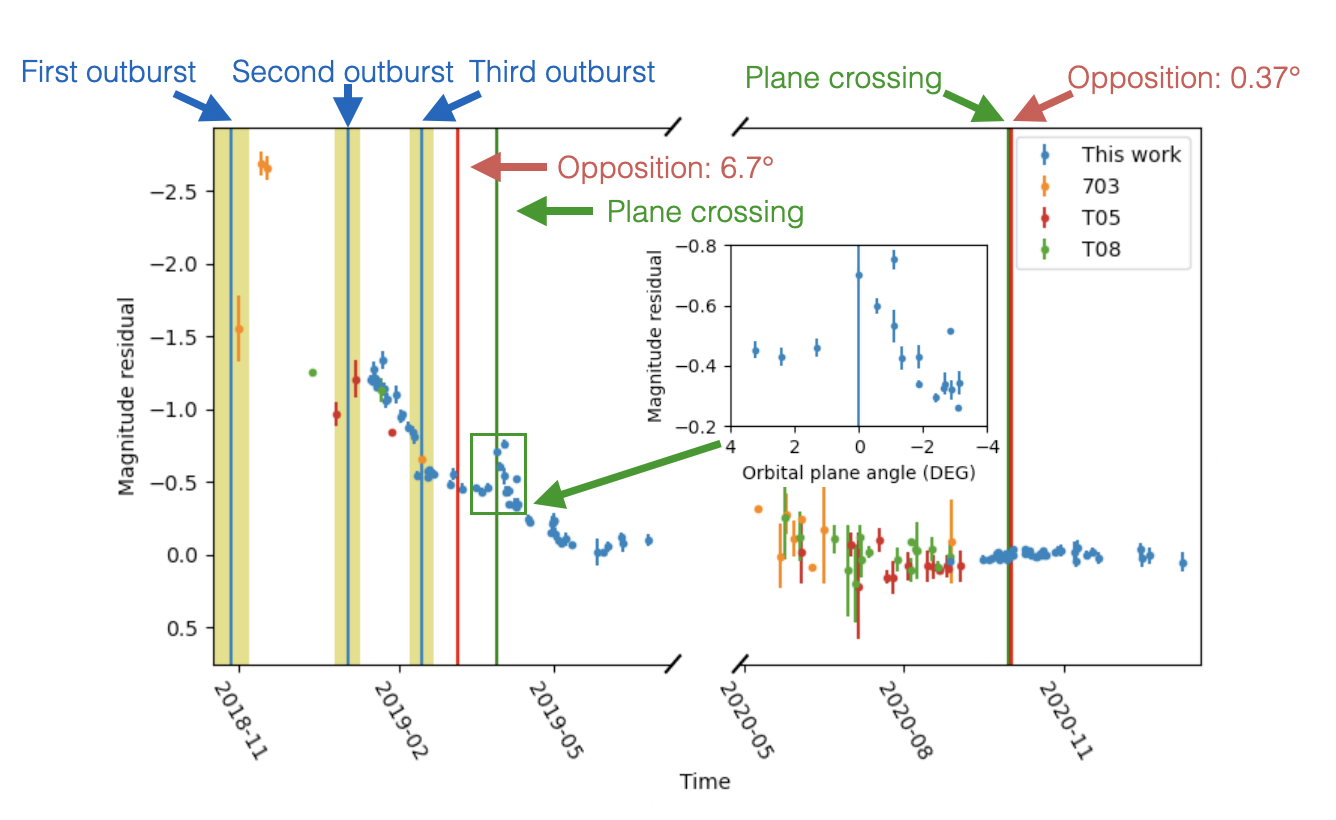}
\caption{Secular variation of Gault's magnitude as a function of time, for the 2019 and 2020 apparition, corrected for distance to the Sun and Earth and  phase curve variation. The blue dots represents the measurements presented in this work. The orange, red and green dots represents data that have been submitted to the Minor Planet Center from respectively the Catalina Sky Survey (CSS; 703) and the ATLAS-HKO (T05) and -MLO (T08) surveys. Times and uncertainties associated to them of the three outburst are shown by blue vertical lines. The times of opposition and their respective phase angle are shown as red vertical line. The times when the Earth crosses the orbital plane of Gault are shown as green vertical line. As expected the 2020 apparition measurements don't show any secular magnitude variation while the 2019 data shows a brightening of more than 2.5 magnitude in late 2018 and did not return back to regular brightness until June 2019. A sharp increase of $\sim$ 0.3 magnitude is observed on 2019 March 25 due to the Earth crossing the orbital plane of Gault.}

\label{fig:JD_Curve}
\end{figure*}

Figure \ref{fig:JD_Curve} represents the magnitude residuals as a function of time for the observations presented in this work, but also for observations that have been submitted to the Minor Planet Center by the Catalina Sky Survey (CSS; 703), and by the Haleakala Observatory station (T05) and Mauna Loa Observatory station (T08) of ATLAS. As our observations are calibrated in the R band, the CSS Gaia G-band observations have been shifted by $-0.225$ mag \citep{Casagrande_2018}, the ATLAS c-band observations by $-0.3$ mag (empirically derived by us to best match other observations), and the ATLAS o-band have been left unchanged.  

In Figure \ref{fig:JD_Curve}, the activity that started right before the first observation of the 2018 apparition is clearly apparent with a 1.5 magnitude brightening of Gault. The second outburst, that occurred in between the first and second observations shown in Figure \ref{fig:JD_Curve} increased the brightness of Gault by approximately another magnitude. At later dates, the dust surrounding Gault slowly dissipates;  Gault returned to its regular magnitude around June 2019. A small, sharp, increase of ~0.3 magnitude is also observed around 2019 March 25. This increase is not due to another activity outburst, but due to the fact that the Earth is crossing the orbital plane of Gault. The dust released during the activity events remains in the orbital plane of their parent body. Thus, when the Earth is crossing this orbital plane, there is an increase of the amount of dust in the line of sight. On the other hand, no variation can be observed during our observation during the 2020 apparition. However, the first data reported to the MPC for the 2020 apparition (around May 2020) seem to display a decrease in flux similar to the one observed after the 2019 activity. Interestingly \citet{Cantelas_2020} reported the observation of two tails on images obtained in July 2020.  

\subsubsection{A YORP driven event?}
\label{sssec:YORP}
In this work we found that the rotation period of Gault is $P = 2.4929$~h and it displayed a lightcurve amplitude of $A = 0.06$ mag. For a strengthless rubble pile asteroids and for a given bulk density ($\rho$) there exist a critical rotation period ($P_{\rm crit}$) for which the centrifugal acceleration equals that of the gravitational acceleration. Here, we made use of the Eq. 1 from \citet{Polishook_2016} to determine the minimum bulk density value needed to hold the asteroid together. We found that for a period of $P = 2.4929$ h and a lightcurve amplitude of 0.06 mag, the minimal bulk density is $\rho = 1.85$ g/cm$^3$.

\citet{Carry_2012} derived the mean density for different taxonomic types. They found that the mean density of S-class asteroids is $\rho = 2.72 \pm 0.54$ g/cm$^3$. Our lower bound for Gault's density is consistent with this mean value.

On the other hand, considering the $1.2465 \pm 0.0003$ h solution for the rotation period we find the bulk density of Gault needs to be at least 7.4 g/cm$^3$. Such a high bulk density value is inconsistent with any reasonable asteroid structure or composition. An asteroid such as Gault rotating at such a speed should disrupt completely while Gault is only showing transient activity outbursts. 

In the previous paragraphs, we considered that Gault is a strengthless body. However, some internal strength could be present. To assess the level of strength that would be needed in order to keep Gault together considering its observed rotation period we apply the Drucker-Prager yield criterion \citep{Holsapple_2007,Polishook_2016}. Details on how to compute this criterion for a specific asteroid can be found in \S4.2 of \citet{Polishook_2016}. Assuming a bulk density of $\rho = 2.5$ g/cm$^3$ we find that the 2.4929 h period solution results in a strengthless body while the 1.2465 h period requires an internal strength of $\sim 1700$ Pa. Such internal strength would imply that Gault is a monolith block. With a diameter of 2.8 km this is very unlikely. It is to be noted that the Drucker-Prager yield criterion requires a bulk density $\rho > 2.25$ in order for Gault to be strengthless.

\subsection{Non-gravitational acceleration}

Gault's episodic activity in 2019, 2016, and 2013 \citep{Chandler_2019} can impart a force that would lead to a modification of its orbit over time. In the case of comets, in many cases, their orbit cannot be modeled using gravitation only and non-gravitational accelerations are routinely needed. However the activity mechanism for Gault is likely different and thus it is unclear if Gault would experience non-gravitational perturbations to its orbit. 

To increase the accuracy of Gault's orbital solution we included the recovered astrometric measurements from the 1958 archival plate and re-measurement of the 1984 observation (\S\ref{sec:astrom}). We also included high accuracy astrometric measurements from our LDT and TRAPPIST observations. 

We calculated the orbit of Gault using two non-gravitational models. The first is the \citet{Marsden_1973} non-gravitational model characterized by three parameters ($A_1$, $A_2$, and $A_3$). The $A_1$, $A_2$ parameters are terms characterizing accelerations in the orbit plane of the object. $A_1$ is defined as the radial term (with respect to the Sun) while $A_2$ is perpendicular to the radial term in the direction of motion of the object. Finally, $A_3$ is a term describing the acceleration normal to the orbit plane of the object. This model is usually used to described the orbit of comets \citep{Krolikowska_2004} or to model the Yarkovsky effect \citep{Farnocchia_2013}. The second model consist of a discrete impulsive $\Delta{\rm v}$ variation at a specified date \citep{Farnocchia_2014}.  

Both models resulted in non-detection of non-gravitational terms within the uncertainties on each individual parameters. These non-detections suggest that Gault's activity does not seem to have net long term effects on its orbit or that they are too small to be detectable with our current knowledge of its orbit. Moreover, we also note that there is no hint of non-gravitational effect in Gault's orbit as it can be properly modeled using gravitation forces only. This is not surprising given the episodic nature of activity displayed by Gault. While sublimation-driven activity tends to be triggered when an active region is illuminated by the Sun, thus producing a net thrust over an extended period of time in a consistent direction. In the case of a YORP induced activity, the mass loss is expected to be short (10-20 days for Gault according to \citet{Jewitt_2019}), stochastic, and with thrusts applied in random directions (mostly due to the rotation of the object).

Using both models simultaneously ($A_1$, $A_2$, $A_3$, and a $\Delta{\rm v}$), we can determine the upper limit on these parameters to be $10^{-11}$, $10^{-13}$, and $10^{-11}$ au/$d^{2}$, and $10^{-6}$ km/s respectively. These upper limits are 100 to 10,000 time smaller than usual values found for comets \citep{Krolikowska_2020} but 10 to 100 time higher than the typical values for the Yarkovsky effect for Near Earth object \citep{Greenberg_2020}. Excluding data prior to 2016, 2012, 2008, 2004, and 2000 also resulted in a non-detection of non-gravitational accelerations. We also tried to apply the $\Delta{\rm v}$ impulse at different dates around the two activity events from late 2018. However, Only one specific date can be used at a time, as each thrust would add an additional degree of freedom that would have for effect to reduce the significance of any detection. None of these attempt resulted in a detection of non-gravitational acceleration. We are thus concluding that if there is orbit modifications, due to the activity events, they are too small to be detectable with our current dataset.

\section{Conclusions}

In this work we presented new observations of the main-belt asteroid (6478)~Gault obtained over two apparitions in 2019 and 2020. We started photometric and spectroscopic observations soon after Gault was found to display activity in December 2018. From the photometry we found that Gault has a rotation period of $P = 2.4929 \pm 0.0003$ h. A second solution corresponding to half that period ($P = 1.2465 \pm 0.0003$~h) cannot be statistically ruled-out, but it would violate the spin-barrier limit for asteroids larger than $\sim 150$~m and seems to be inconsistent with a subset of our data (Figure \ref{fig:Gault_Period_Last}). Our preferred rotation period solution of $P = 2.4929 \pm 0.0003$~h is consistent with an asteroid undergoing YORP spin-up and mass loss of a moderately elongated asteroid (lightcurve amplitude of 0.06 mag) with a bulk density of 1.85 g/cm$^3$. 

The spectroscopic observations confirm that Gault can be classified either as an S or a Q-type object in the Bus-DeMeo taxonomic system \citep{Demeo_2009}. The Gault spectrum seems to be intermediate to S and the Q-types, but we suggest that it is closer to a low slope S-type rather than a Q-type. As large scale (global) activity might be expected to turn a weathered S-type surface into a fresh Q-type, we suggest that Gault's activity in 2018/2019 may have been localized to region(s) close to the equator and that its overall surface still appears weathered.

We monitored the photometry of Gault throughout the 2019 and 2020 apparition and constructed a phase curve using the 2020 data. We modelled this phase curve using the $H$, $G1$, $G2$ formalism and found $H = 14.81 \pm 0.04$, $G1=0.25 \pm 0.07$, and $G2= 0.38 \pm 0.04$. These values are consistent with an S or Q-type asteroid. Using the relation linking the $G1$ and $G2$ parameters to the albedo we inferred its albedo to be $p_{\rm v} = 0.26 \pm 0.05$. Finally, combining our new determination of $H$ and  $p_{\rm v}$, we find that Gault has a diameter $D=2.8^{+0.4}_{-0.2}$ km. Analysis of the photometric phase curve with Hapke radiative transfer models shows that the surface of Gault is composed of grain with size ranging from 100-500 $\rm{\mu}$m. The photometry from the 2019 apparition clearly shows the sign of activity while such signature is absent in 2020. Deep stacks on the asteroids did not reveal any tail or coma and comparisons between the PSF profile of Gault and reference stars also resulted in negative results. We are thus concluding that Gault did not experience any detectable activity during our observations.

Finally, we conducted a search for archival observations of Gault and found new detections on plates dating back to 1958. These new astrometric measurements increase the orbital arc by a factor of 1.7. We used these new detection along with our high accuracy astrometric measurements from 2020 to analyse the orbit of Gault. We fit the orbit with two different non-gravitational models, first the \citet{Marsden_1973} non-gravitational model and a $\Delta{\rm v}$ impulse on 2018 November 5 to coincide with the inferred onset of the most recent episode of activity. Both models resulted in a non-detection of non-gravitational effects. The addition of non-gravitational forces does not provide significant improvement of the overall fit of the orbit. 
We interpret this result as attributable to YORP activity that pushes Gault in random directions resulting in a net zero non-gravitational effect or the activity level to be too low to significantly affect the orbit. 
However, it remains possible that future outbursts could be detected through instantaneous modification of the orbit.

\section*{Acknowledgements}

M.D., N.M, B.S., and A.G. acknowledge funding from NASA NEOO grant NNX17AH06G in support of the Mission Accessible Near-Earth Object Survey (MANOS).

M.D. and N.M acknowledge funding from NASA NEOO grant 80NSSC21K0045 in support of the Second Lunation NEO Follow-up.

TRAPPIST is a project funded by the Belgian Fonds (National) de la Recherche Scientifique (F.R.S.-FNRS) under grant FRFC 2.5.594.09.F. TRAPPIST-North is a project funded by the University of Li\`{e}ge, in collaboration with the Cadi Ayyad University of Marrakech (Morocco). E. Jehin is a FNRS Senior Research Associate. 

Part of this research was conducted at the Jet Propulsion Laboratory, California Institute of Technology, under a contract with the National Aeronautics and Space Administration (80NM0018D0004).

These results made use of the Lowell Discovery Telescope (LDT) at Lowell Observatory. Lowell is a private, non-profit institution dedicated to astrophysical research and public appreciation of astronomy and operates the LDT in partnership with Boston University, the University of Maryland, the University of Toledo, Northern Arizona University and Yale University. The Large Monolithic Imager (LMI) was built by Lowell Observatory using funds provided by the National Science Foundation (AST-1005313). The upgrade of the DeVeny optical spectrograph has been funded by a generous grant from John and Ginger Giovale and by a grant from the Mt. Cuba Astronomical Foundation. 

Part of the photometric data in this work were obtained at the C2PU facility (Calern Observatory, O.C.A.).

The NTT data presented in this work have been obtained through the ESO programme ID 0103.C-0224. 

J.d.W. and MIT gratefully acknowledge financial support from the Heising-Simons Foundation, Dr. and Mrs. Colin Masson and Dr. Peter A. Gilman for Artemis, the first telescope of the SPECULOOS network situated in Tenerife, Spain.

The ULiege's contribution to SPECULOOS has received funding from the European Research Council under the European Union's Seventh Framework Programme (FP/2007-2013) (grant Agreement n$^\circ$ 336480/SPECULOOS), from the Balzan Prize Foundation, from the Belgian Scientific Research Foundation (F.R.S.-FNRS; grant n$^\circ$ T.0109.20), from the University of Liege, and from the ARC grant for Concerted Research Actions financed by the Wallonia-Brussels Federation. MG is F.R.S-FNRS Senior Research Associate. 

SFG acknowledges funding from UK STFC (grants ST/P000657/1 and ST/T000228/1).

The Isaac Newton Telescope is operated on the island of La Palma by the Isaac Newton Group of Telescopes in the Spanish Observatorio del Roque de los Muchachos of the Instituto de Astrofísica de Canarias.

This material is based upon work supported by the National Science Foundation Graduate Research Fellowship Program under grant No.\ 2018258765. Any opinions, findings, and conclusions or recommendations expressed in this material are those of the author(s) and do not necessarily reflect the views of the National Science Foundation. Computational analyses were carried out on Northern Arizona University's Monsoon computing cluster, funded by Arizona's Technology and Research Initiative Fund. This work was made possible in part through the State of Arizona Technology and Research Initiative Program.

This work made use of the \textit{astropy} software package \citep{AstroPy2013}. This research has made use of the The Institut de M\'ecanique C\'eleste et de Calcul des \'Eph\'em\'erides (IMCCE) SkyBoT Virtual Observatory tool \citep{Berthier:2006tn}. This research has made use of data and/or services provided by the International Astronomical Union's Minor Planet Center. This research has made use of NASA's Astrophysics Data System. This work made us of the Vizier catalog service \citep{Ochsenbein2000Vizier}, the Source Extractor \citep{Bertin:1996hfa} and Scamp software package \citep{Bertin2006SCAMP}, the Sloan Digital Sky Survey (SDSS) Data Release 9 \citep{SDSS2012DR9} catalog, the Gaia Data Release 2 catalog \citep{Gaia2018DR2}, and the Aperture Photometry Tool software tool \citep{Laher2012APT}. Image stacking was performed in part with the Siril software tool (\url{https://www.siril.org}) which acknowledges \cite{huber2009robust} and Juan Conejero of Pixinsight \url{http://www.pixinsight.com}. This research has made use of SAO Image DS9, developed by Smithsonian Astrophysical Observatory \citep{2003ASPC..295..489J}. Astrometry was aided by the AstrometryNet software and service (\url{https://www.astrometry.net}; \citealt{lang2010astrometrynet}).

\section*{Data Availability}

All photometric and spectroscopic are available on the Centre de donn\'{e}es astronomiques de Strasbourg (CDS). Raw data from the NTT and both TRAPPIST will be available on the ESO archive after the respectively a 12 and 24 months priority period from their acquisition. The POSS I 103aO plate ID 1619 is available in the supplementary electronic material of this paper. The SERC-EJ plate ID 9004 can be retrieved here: \url{https://archive.stsci.edu/cgi-bin/dss\_plate\_finder}. All additional information or raw data product will be made available upon reasonable request.



\bibliographystyle{mnras}
\bibliography{Gault_Paper.bib} 




\appendix

\onecolumn

\section{Summary of photometric observations}

\label{ap:table}
Table \ref{Tab:Phot_Data} list all the photometric observations obtained during the 2019 and 2020 apparition of Gault that were used in this work. The full table is available online

\begin{table*}
\caption{Summary of Gault photometric observations. The full table is available online}
\label{Tab:Phot_Data}
\begin{tabular}{ccccccccccc}
\hline
Date & Julian date &Filter & \# of images & V mag  & Telescope & $\lambda$ & $\beta$ & $r$  & $\Delta$ & $\alpha$ \\
 & & & & (mag)      &  &(deg)         & (deg)  & (au)  & (au) &($^{\circ}$) \\
 \hline
2019-01-11 & 2458494.63459 & r & 59 & 18.6 & C2PU & 168.85 & -18.68 & 2.464  & 1.821 & 20.2 \\
2019-01-12 &2458495.64507 & r & 50 & 18.5 & C2PU & 168.86 & -18.70 & 2.462  & 1.808 & 20.1 \\
2019-01-13 & 2458496.76228 & B & 2 & 18.5 & TS & 168.86 & -18.72 & 2.460  & 1.794 & 19.9 \\
2019-01-13 & 2458496.76501 & Rc & 119 & 18.5 & TS & 168.86 & -18.72 & 2.460  & 1.794 & 19.9 \\
2019-01-13 & 2458496.76537 &V & 2 & 18.5 & TS & 168.86 & -18.72 & 2.460  & 1.794 & 19.9 \\
\hline
\end{tabular}
\end{table*}


\bsp	
\label{lastpage}
\end{document}